\documentclass[12pt,preprint]{aastex}

\shorttitle{Ultracompact Binaries in Distant  Globular Clusters}
\shortauthors{Bildsten and Deloye}

\begin{document}

\title{Ultracompact Binaries as Bright X-Ray Sources in Elliptical Galaxies}

\author{Lars Bildsten\altaffilmark{1,2} and Christopher
J. Deloye\altaffilmark{2}}
\altaffiltext{1}{Kavli Institute for Theoretical Physics, 
Kohn Hall, University of California, Santa Barbara, CA 93106;
bildsten@kitp.ucsb.edu}
\altaffiltext{2}{Department of Physics, 
Broida Hall, University of California, Santa Barbara, CA 93106;
cjdeloye@physics.ucsb.edu}

\begin{abstract}
 {\it Chandra} observations of distant ($D\sim 10$ Mpc) elliptical
galaxies have revealed large numbers of Low Mass X-ray Binaries
(LMXBs) accreting at $\dot M > 10^{-8} M_\odot$ yr$^{-1}$.  The
majority of these LMXBs reside in globular clusters (GCs) and it has
been suggested that many of the field LMXBs also originated in GCs. We
show here that ultracompact binaries with orbital periods of 8-10
minutes and He or C/O donors of $0.06-0.08M_\odot$ naturally provide
the observed $\dot M$'s from gravitational radiation losses alone.
Such systems are predicted to be formed in the dense GC environment, a
hypothesis supported by the 11.4 minute binary 4U 1820-30, the
brightest persistent LMXB in a Galactic GC.  These binaries have short
enough lifetimes ($<3\times 10^6$ years) while bright ($L>10^{38} {\rm
erg \ s^{-1}}$) that we calculate their luminosity function under a
steady-state approximation. This yields a luminosity function slope in
agreement with that observed for $6\times 10^{37}{\rm erg \
s^{-1}}<L<5\times 10^{38} {\rm erg \ s^{-1}}$, encouraging us to use
the observed numbers of LMXBs per GC mass to calculate the accumulated
number of ultracompact binaries.  For a constant birthrate over 8
Gyrs, the number of ultracompact binaries which have evolved through
this bright phase is $\sim 4000$ in a $10^7M_\odot$ GC, consistent
with dynamical interaction calculations. Perhaps most importantly, if
all ultracompacts become millisecond radio pulsars, then the observed
normalization agrees with the inferred number of millisecond radio
pulsars in 47 Tuc and Galactic GCs in general. 

\end{abstract}


\section{Introduction}

 The {\it Einstein} observatory allowed for the discovery and study of
  X-ray emission from elliptical and S0 galaxies (e.g. Forman, Jones
  \& Tucker 1985), finding that the emission likely had two distinct
  contributors (Trinchieri \& Fabbiano 1985, Canizares, Fabbiano, \&
  Trinchieri 1987).  The softer emission is from hot interstellar gas
  (Forman et al 1985, Trinchieri, Fabbiano, \& Canizares 1986) and
  dominates the total X-ray emission from the most massive
  ellipticals.  Observations with ASCA (Matsumoto et al. 1997) showed
  that a harder (5-10 keV) spectral component scaled linearly with the
  bolometric luminosity of the galaxy, confirming its suggested origin
  as the unresolved contributions of numerous Low Mass X-Ray Binaries
  (LMXBs; Trinchieri \& Fabbiano 1985; White, Nagase, \& Parmar 1995;
  Irwin \& Sarazin 1998).

 The exquisite point source sensitivity of {\it Chandra} led to the
 discoveries of the expected LMXBs (e.g. Sarazin, Irwin \& Bregman
 2000; Blanton, Sarazin \& Irwin 2000; Sarazin, Irwin, \& Bregman
 2001; Finoguenov \& Jones 2001; Angelini, Loewenstein, \& Mushotzky
 2001; Kraft et al. 2001; Kundu, Maccarone, \& Zepf 2002; Di Stefano
 et al. 2003; Randall, Sarazin \& Irwin 2004).  These observations
 (see Gilfanov 2004 and Kim \& Fabbiano 2004) typically find 20-50
 LMXBs with $L> 10^{38} {\rm erg \ s^{-1}}$ from elliptical galaxies
 with stellar masses of $(1-3)\times 10^{11}M_\odot$.  This has
 provided a large sample of very bright LMXBs that is unattainable
 from the Milky Way or even M31, allowing for a new probe of binary
 evolution.  Gilfanov (2004) showed that the LMXBs of these elliptical
 and S0 galaxies are consistent with a single luminosity function
\begin{equation} 
{dN\over dL}\propto {1\over L^{\alpha}}, 
\end{equation}
from $10^{37} {\rm erg \ s^{-1}}$ to a break value
$L_B=5^{+1.4}_{-0.7}\times 10^{38} {\rm erg \ s^{-1}}$.  with
$\alpha=1.64\pm 0.22$.  Kim and Fabbiano (2004) recently found a
similar result for $L>6\times 10^{37} {\rm erg \ s^{-1}}$ of
$\alpha=1.80\pm 0.2$ and $L_B=4.8\pm 1.2 \times 10^{38} {\rm erg \
s^{-1}}$.  For $L>L_B$, the luminosity function steepens to
$\alpha=2.7\pm 0.5$.  The total number of X-ray sources scales with
the galactic mass, with $\approx 20$ sources of $L>10^{38} {\rm erg \
s^{-1}}$ per $10^{11}M_\odot$ (Gilfanov 2004), or a total X-ray
luminosity that scales linearly with the $K$ band luminosity as
$L({\rm LMXB})=1.5\pm 0.6 \times 10^{40} {\rm erg \ s^{-1}}$ for
$L_K=10^{11}L_{K,\odot}$ (Kim \& Fabbiano 2004).
 
  Starting with the original observation of NGC 4697 (Sarazin et
  al. 2001), it has become clear that many (20-70\%) of these LMXBs
  are residing in globular clusters (Angelini et al. 2001; Kundu et
  al. 2002; Maccarone, Kundu \& Zepf 2003; Minnitti et al. 2004).
  Kundu et al. (2002) showed that there was no difference in $dN/dL$
  for LMXBs in and out of GCs (both had $\alpha=1.55\pm 0.15$ and
  $L_B\approx 3\times 10^{38} {\rm erg \ s^{-1}}$) in NGC 4472 ,
  supporting the suggestion of White, Sarazin and Kulkarni (2002) that
  most of the bright LMXBs in ellipticals are made in GCs. Kundu et
  al. (2002) showed that nearly 4\% of NGC 4472 GCs host a bright
  LMXB, with metal rich clusters about 3 times more likely to host an
  LMXB (confirmed in NGC 4365 and NGC 3115 by Kundu
  et. al. 2003). Sarazin et al. (2003) studied four galaxies and found
  that the specific incidence of LMXBs in GCs is about one source with
  $L>10^{38} {\rm erg \ s^{-1}}$ per $10^7 L_{\odot,I}$, so that a
  star in a GC is about 1000 times more likely to be a donor than a
  star in the field.  Such a large enhancement of the incidence of
  LMXBs in GCs was first found in the Milky Way (Katz 1975; Clark
  1975) and is indicative of the important role of interactions in
  creating mass transferring binaries in GCs (see Hut et al. 1992 for
  an overview).

  Though numerous LMXBs have been found, the nature of the donor star
in this old stellar population remains a mystery. Piro \& Bildsten
(2002) showed that for field LMXBs, the simplest way to reach
$10^{38} {\rm erg \ s^{-1}}$ (or $\dot M=LR/GM\approx 10^{-8} M_\odot$
yr$^{-1}$  for accretion onto a neutron star of $M=1.4M_\odot$ and 
$R=10 {\rm \  km}$) is to have a red giant branch
 star fill the Roche lobe with orbital periods of days or longer. They
 noted that these wide binaries are nearly always transient accretors,
 and that multiple {\it Chandra} observations would easily identify
 them (e.g. Kraft et al. 2001). However, calculating the resulting
 $dN/dL$ is impossible given our poor state of knowledge of the
 transient duty cycle.

We argue here that the most likely type of mass transferring binary
responsible for the bright end of the luminosity function in distant
elliptical galaxies (especially those in GCs) are ``ultracompact''
systems consisting of either a He or C/O white dwarf donor of mass 
$M_c\approx 0.04-0.08M_\odot$ which is filling its Roche lobe in a
5-10 minute orbit with a neutron star. In \S 2, we summarize the 
observational clues that led us to this conjecture and
derive the resulting $dN/dL$, showing that the observed power law is
naturally explained. In \S 3, we use the observed normalizations from
{\it Chandra} to compare to the predicted formation rate of
ultracompacts in GCs from dynamical interactions. We also predict the
expected number of millisecond radio pulsars (MSPs) assuming that all
ultracompacts become MSPs. We close in \S 4 by noting more 
observational tests, and discussing future work.
 
\section{Ultracompact Binary Formation and  Evolution} 

  There are a number of facts that support our conjecture. The first
is that the brightest LMXB in a galactic GC (see Table 1 in Deutsch,
Margon \& Anderson 2000) and the only one with a luminosity comparable
to that detected in a distant elliptical is the $L\approx (4-7) \times
10^{37} {\rm erg \ s^{-1}}$ source 4U~1820-30 in NGC 6624. Stella,
Priedhorsky \& White (1987) found a coherent 11.4 minute periodicity
in 4U1820-30 with a peak to peak variation in the X-rays of $\approx
3$\%, which is now securely identified as the orbital period (van der
Klis et al. 1993).  The observed properties of the thermonuclear
flashes from this system are best explained by accretion of nearly
pure helium (e.g.  Bildsten 1995; Strohmayer \& Brown 2002; Cumming
2003). The second clue is that the observed luminosity break, $L_B$,
is much closer to the Eddington limit of a neutron star (NS) 
accreting material with two baryons per electron (e.g. He or C/O)
\begin{equation} 
L_{\rm Edd}={8\pi G M m_p c\over \sigma_{\rm Th}} = 3.5\times 10^{38}
{\rm erg \ s^{-1}}\left(M\over 1.4 M_\odot\right),
\end{equation}
(where $\sigma_{\rm Th}$ is the Thomson scattering cross section) 
than accretion of cosmic composition  This has been noted previously (e.g. Kim and Fabbiano 2004) but none made the connection to ultracompacts. 

The last fact is that ultracompact progenitors (i.e. a NS and white
dwarf in a few hour orbit) are vigorously produced in GCs (Verbunt
1987; Davies 1995; Rasio, Pfahl and Rappaport 2000). Verbunt (1987)
noted that one way to form these systems is a direct stellar collision
of a NS with a red giant that would trigger a common envelope (CE)
event (though simulations of Rasio and Shapiro 1991 found no common
envelope). The CE would allow for inspiral of the NS and red giant He
core (of mass $M_{c,i}\approx 0.1-0.4M_\odot$) to short enough orbital
periods that gravitational wave losses drive it into contact within a
Hubble time and initiate mass transfer.  Rasio et al. (2000) showed
that a much more likely formation channel is an exchange interaction
of the NS with a primordial binary (e.g. Hut, Murphy \& Verbunt 1991;
Sigurdsson \& Phinney 1993) when the GC was young. The higher main
sequence turnoff mass would then trigger unstable mass transfer and a
CE event and inspiral of the He core when the star leaves
the main sequence.  Some systems formed this way could also be ejected
from the GC and appear as LMXBS at a later date when the gravitational
radiation losses have driven them into contact.
 
  At these very short orbital periods (1-10 minutes), $\dot M$ is set
by the rate of angular momentum loss from gravitational radiation
\begin{equation} 
{\dot J\over J}= -{32 G^3 M M_c(M+M_c)\over 5c^5 a^4}, 
\end{equation}
where $a$ is the orbital separation.  Under the constraint of
conservative mass transfer and Roche lobe filling, $R_c=0.46
a(M_c/(M+M_c))^{1/3}$, $\dot M$ relates to the angular momentum loss
rate as
\begin{equation} 
-{\dot J\over J}=-{\dot M_c\over M_c}\left({5\over 6}+{n\over 2}-{M_c\over M}\right),
\end{equation} 
where $n=d\log R_c/d\log M_c$ is the rate of change of the WD radius
under mass loss. Bildsten (2002) and Deloye and Bildsten (2003;
hereafter DB) calculated the impact of finite entropy on the value of
$n$, which is most pronounced at longer orbital periods where the
donor is even less massive, $\approx 0.01 M_\odot$.  However, the
composition of the WD does matter, as Coulomb physics makes the C or O
donors more compact than He donors at a given mass and modifies $n$,
allowing it to range from $n=-0.1$ to $n=-0.3$ (see Figure 1).
 
Figure 1 shows the relations between $M_c$, $n$ and orbital period as
a function of the accretion luminosity, $L$, found by integrating
equations (3) and (4) with the WD models of DB and an initial NS mass
of $M=1.4M_\odot$. This shows that the companion masses required
($M_c=0.04-0.08M_\odot$) to reach $3\times 10^{-8} M_\odot {\rm
yr^{-1}}$ are consistent with the evolution of a system with an
initial WD donor mass of $>0.1M_\odot$.  We plot the evolution here
during the $L<L_{\rm Edd}$ stage and attribute the observed cutoff in
$dN/dL$ to either the impact of super-Eddington accretion or (as noted
below) a predominance of ultracompact progenitors with $0.10-0.12
M_\odot$ He cores.  We see no reason to attribute the $L>L_{\rm Edd}$
sources to accretion onto black holes and think it most likely that
these sources are ultracompacts accreting via a disk at a super-Eddington
rate. {\it The best confirmation of our conjecture would be the
discovery of the orbital periodicity in the range of Figure 1.}

Ultracompacts evolve through the observable $L$ range in $<10^7$
years, a timescale much shorter than either the age of the GCs or any
reasonable expectations of a timescale over which the birthrate of
mass-transferring systems might change. Hence, their total number is
indicative of their current birthrate and we can demand continuity
within the accreting population to directly calculate $dN/dL$.  We
start by combining the evolution equations to yield a simple analytic
form for $\dot M_c\propto -M_c^\beta$ where $\beta=10/3-4n$ in the
limit of $M_c\ll M$.  For $n=-0.2$, this gives $\dot M_c\propto
-M_c^{4.13}$. Integrating this equation (and setting $n$ and $M$
constant) yields $M_c$ as a function of time, $t$, since the onset of
mass transfer, $t\propto M_c^{1-\beta}(1-(M_{c,i}/M_c)^{1-\beta})$,
where $M_{c,i}$ is the initial WD donor mass set by the progenitor
scenario, which can range from $\approx 0.1-0.4M_\odot$ for He cores
to $\approx 0.6M_\odot$ for C/O cores.

 In the limit that $M_c\ll M_{c,i}$, we get the simple relation $\dot
M\propto (1/t)^{\beta/(\beta-1)}$. The amount of time the binary
spends with a luminosity in excess of $L$ is $t\propto
(1/L)^{(\beta-1)/\beta}$, which immediately yields $N(>L)\propto t$. 
The resulting value of $\alpha$ (in equation 1) is then 
\begin{equation}
\alpha={{\beta-1}\over \beta} +1={{17-24n}\over {10-12n}},
\end{equation}
or $\alpha=1.77$ for $n=-0.25$, in excellent agreement with
the observed values of $\alpha=1.64\pm 0.22$ (Gilfanov 2004) and
$\alpha=1.8\pm 0.2$ (Kim and Fabbiano 2004). Clearly, there is little
difference in $\alpha$ for the range of $n$ expected from varied
compositions. The power law steepens
 for $M_c$ close to $M_{c,i}$, which could explain the observed break
 at $L_B$ if nearly all of the ultracompact progenitors had
 $M_{c,i}\approx 0.10-0.12 M_\odot$. This mass is the He
 core mass at the turnoff, but we know of no dynamical calculations
 that point to such a narrow $M_{c,i}$ distribution.
  
   Most importantly, we have shown that the luminosity function slope
 is nearly independent of both the age of the GC and the type of
 initial WD donor (i.e. He or C/O). This universality and remarkable
 agreement with the observations allows us to use the {\it Chandra}
 derived incidence of LMXBs to calculate the ultracompact birthrates
 and the total number of such systems made over a Hubble time.
 
\section{Binary Interactions and Millisecond Radio Pulsars in Globular Clusters}
  
  The identification of elliptical LMXBs with ultracompact binaries
has broad repercussions.  The simplest one to state is that the much
smaller donor mass (compared to main sequence donors) naturally
shortens the LMXB lifetimes and dramatically increases the implied
LMXB birthrate. In their study of four elliptical galaxies, Sarazin et
al (2003) found a specific LMXB incidence within GCs of $\approx 1 $
source with $L>10^{38} {\rm erg \ s^{-1}}$ per $10^7 L_{\odot,I}$,
consistent with the $\approx 10^{-7}/M_\odot$ found by Kundu et
al. (2003). This incidence is independent of GC mass, but dependent on
GC metallicity (Kundu et al. 2003). An ultracompact at $L=10^{38} {\rm
erg \ s^{-1}}$ has $\dot M\approx 8.4\times 10^{-9} M_\odot {\rm
yr^{-1}}$, a He donor of $M_c\approx 0.06M_\odot$ and an age of
$2\times 10^6$ years, implying an {\it ultracompact birthrate of one
new bright LMXB every $2\times 10^6$ years in a $10^7M_\odot$ GC. }

 Kundu et al.'s (2003) preliminary study found that the incidence of
bright LMXBs is consistent with being independent of GC age over the
range of 3-11 Gyr, allowing us to integrate the birthrate over 8 Gyr
to obtain $\sim 4000$ WD-NS binaries capable of undergoing stable mass
transfer in a $10^7M_\odot$ cluster. Davies' (1995) calculations found
5-50 WD-NS binaries that will make contact within 10 Gyrs per 1000
initial binaries. If the initial binary fraction was $10\%$, then the
observed rate is consistent with Davies (1995) lower estimate. Ivanova
and Rasio (2004) found $\sim 30$ WD-NS capable of making contact (with
a roughly constant rate in time, see Rappaport et. al. 2001 for a
declining rate in time) during the evolution of a cluster model with a
core density of $10^5 M_\odot {\rm pc^{-3}}$ (similar to 47 Tuc) and a
total mass after a Hubble time of $2\times 10^5 M_\odot$; a factor of
three less than implied by the {\it Chandra} observations and our
identification with ultracompacts. Future comparisons must take into
account the initial WD mass and its impact on the binary's fate 
once contact is established (i.e. merging versus stable mass
transfer).

  We can also check our hypothesis by comparing these total numbers to
the supposed endpoints of the mass transfer, a recycled MSP 
either in a tight binary (Rasio et al. 2000) or
isolated (see Bildsten's (2002) discussion of the likelihood of
this). Kulkarni, Narayan and Romani (1990) calculated that the MSP
birthrate in all galactic GCs ($\sim 10^8 M_\odot$) was $\sim 1$ every
$10^6$ years for a total number of GC MSPs of $\sim 10^4$.  Assuming
that the X-ray lifetimes of the GC LMXBS was $10^9$ years (consistent
with mass transfer on the main sequence, but likely inconsistent with
the lifetime implied by the calculated formation rate; Hut, Murphy \&
Verbunt 1991) and using the $\sim 10$ persistent sources, they found a
GC LMXB birthrate of one every $10^8$ years.  They noted that a factor
of 100 reduction in the X-ray lifetime was needed to alleviate the
discrepancy with the MSP birthrate. In retrospect, it is clear that
the GC LMXB birthrate is completely dominated by the highest $\dot M$
source, the ultracompact 4U~1820-30 with an age of $2.5\times 10^6$
years. Though the numbers are small, its mere existence gives a GC
LMXB birthrate a factor of 100 higher. Rather than depend on
4U~1820-30 to derive the rate, if we use the rate found in elliptical
GCs, then there should be $\approx 40,000$ MSPs in all of the
galactic GC's, consistent with Kulkarni et al.'s (1990) estimate. 

 Finally, we compare the expectations from the ultracompact
formation rate with the MSP population in the galactic GC 47 Tuc.  The
discovery of 20 MSPs in 47 Tuc (Camilo et al. 2000) led Camilo et
al. (2000) to speculate that there might be as many as 200 MSPs in
this $10^6M_\odot$ cluster. Later X-ray observations (Grindlay et
al. 2002; Edmonds et al. 2003) bring this number down to 100
(Camilo, F., priv. comm.). The implied rate from ellipticals
would yield $\approx 400$ MSPs, just slightly high. 
 
\section{Summary and Future Work}

 Our identification of these distant LMXBs with ultracompact binaries
naturally explains the observed luminosity function from {\it Chandra}
observations and yields the LMXB birthrate of one new mass
transferring binary every $2\times 10^6$ years per $10^7M_\odot$ of
GCs. Rather remarkably, this derived birthrate from distant
ellipticals agrees with both dynamical calculations (e.g. Ivanova \&
Rasio 2004) and the observed number of MSPs in galactic GCs,
especially 47 Tuc. The much lower donor mass ($\approx 0.06M_\odot$)
has alleviated the ``birthrate'' problem often discussed for LMXBs and
MSPs in galactic GCs (e.g. Kulkarni et al. 1990).

  The simplest way to prove this hypothesis is to find the 5-10 minute
orbital periods. If we use 4U~1820-30 as our example, the level of
orbital variability in the X-rays could be as low as a few percent,
allowing for {\it Chandra} searches amongst the few bright LMXBs in
M31 GCs while {\it XMM-Newton } could probe much deeper. The source
4U~1820-30 cycles in luminosity by about a factor of 3 over a 171 day
cycle for unknown reasons (Chou \& Grindlay 2001).  Such variability
is easily detected by {\it Chandra}, and indeed variability at this
level has been reported for bright LMXBs in M31 (e.g. Trudolyubov \&
Priedhorsky 2004). However, attributing such behavior as unique to
ultracompacts is much harder. Ultraviolet observations of 4U~1820-30
(King et al 1993; Arons and King 1993) confirmed the 11.4 minute orbit
and similar work could be done with {\it HST} amongst the bright
systems in M31.

 The lower luminosities that are visible in the GC sources in our
galaxy and in M31 have yet to be probed by {\it Chandra} in distant
ellipticals. For ultracompact binaries, the expectation is that our
derived $dN/dL$ will continue until a $L$ is reached where the systems
become transients (presuming no dramatic episode associated with MSP
turn-on). DB showed that in the absence of X-ray heating in the outer
disk, this would occur at $L\sim 10^{37} {\rm erg \ s^{-1}}$, whereas
with X-ray heating, the disks can remain stable down to much lower
X-ray luminosities. Unfortunately, the current populations of
ultracompacts in our galaxy don't provide a stringent test of which
case is correct. {\it Chandra} may be able to identify this cutoff 
with deeper observations of nearby ellipticals. 

 There is still much to explain. Clearly, not all bright LMXBs are
ultracompact binaries, both because the MSPs in wide binaries with He
WDs (orbital periods longer than a day) cannot be made from
ultracompacts and because roughly half the galactic GC sources are
clearly hydrogen accretors (Kuulkers et al. 2003). However, the strong
expectation is that most of the H donors with $L$ large enough to detect
at 10 Mpc are transient accretors (e.g. Piro \& Bildsten 2002), making
secure predictions of $dN/dL$ difficult. Observations of the bright
LMXBs in M31 (Di Stefano et al. 2002; Trudolyubov \& Priedhorsky 2004)
have motivated a stable mass transfer scenario involving thermal
timescale mass transfer from evolved stars (Di Stefano et al. 2002),
but $dN/dL$ was not derived. We have also not explained the higher
incidence of bright LMXBs in metal-rich GCs (Kundu et al. 2002;
Maccarone et al. 2004).

\acknowledgments
 
  We thank F. Camilo, D. Chakrabarty, D.  Fox, V.  Kalogera, T.
 Maccarone, S. Phinney, F. Rasio, R. Rutledge,  and S. Zepf for
 discussions. 
L.B. thanks the Center for Gravitational Wave Physics at
 Penn State (supported by the NSF under cooperative agreement
 PHY01-14375) for organizing a workshop that triggered this idea,
 Caltech for hospitality during the writing of the article, and
 S. Phinney for many helpful discussions.  
This work was supported by the
 NSF under grants PHY99-07949 and AST02-05956.

\clearpage

\begin{figure}
\plotone{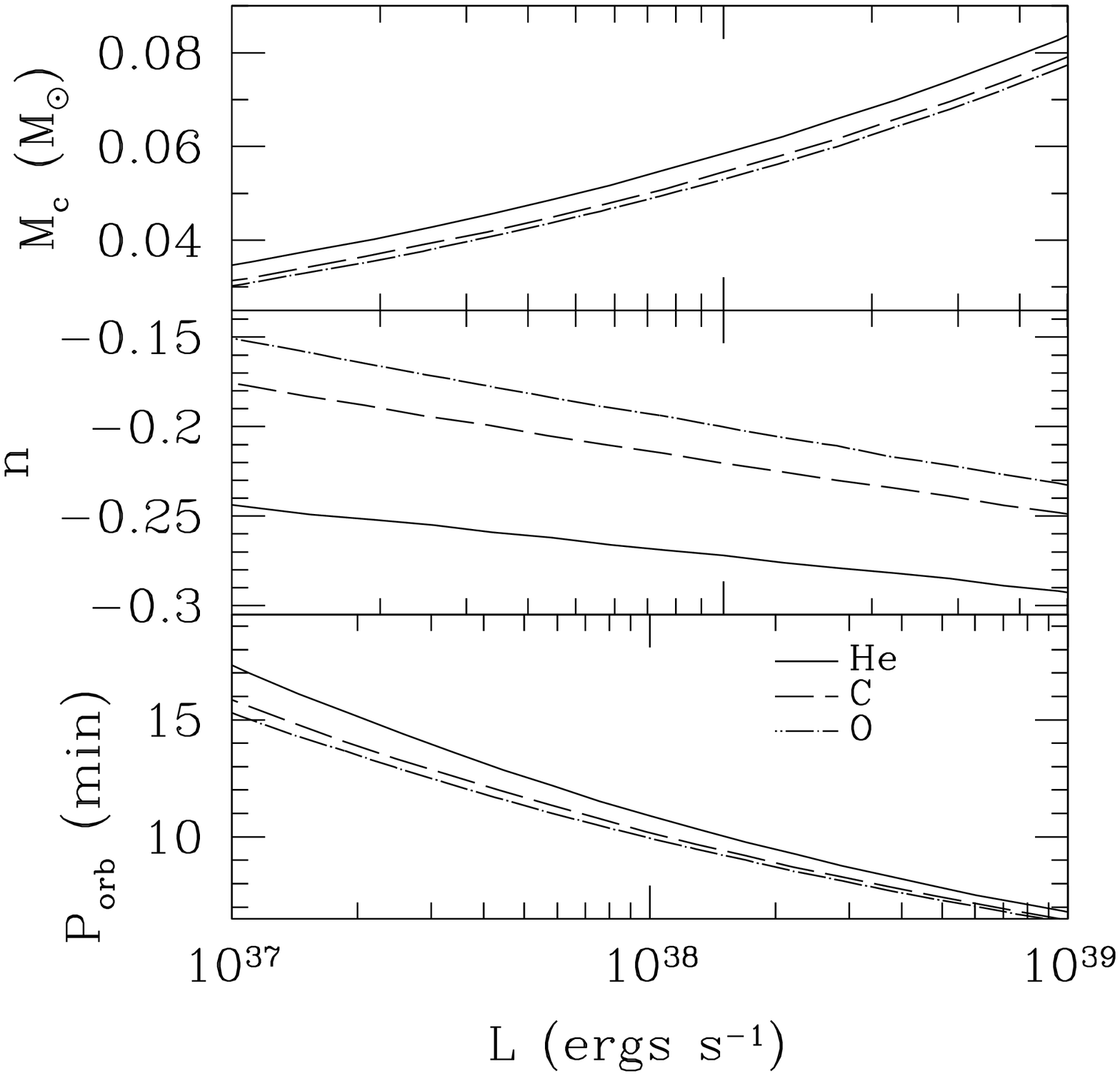} 
\figcaption{ The relation between donor mass, $n$, 
and orbital period as a function of $L$ for cold donors of different
compositions (DB) and a NS mass of $1.4M_\odot$. We neglected entropy
effects since those modifications are small compared to those from
composition differences, which are clearly important. The solid line
is for a Helium  donor, whereas the dashed is for Carbon and dot-dashed for
Oxygen.}
\end{figure}

\end{document}